\documentclass[sigconf]{acmart}

\usepackage{color}

\newcommand{\heading}[1]{\vspace*{0.5mm}\noindent\textbf{#1.}}
\AtBeginDocument{%
  \providecommand\BibTeX{{%
    \normalfont B\kern-0.5em{\scshape i\kern-0.25em b}\kern-0.8em\TeX}}}

\usepackage{bbm}
\usepackage{multirow}
\usepackage[inline]{enumitem}
\usepackage{graphicx}
\usepackage{sistyle}
\usepackage[ruled]{algorithm2e}
\usepackage{booktabs}
\usepackage{caption}
\SIthousandsep{,}
\usepackage{makecell}
\usepackage[skip=0pt]{caption}
\usepackage{amsmath}
\usepackage{subfigure}
\usepackage{hyperref}
\usepackage{array} 
\usepackage{graphicx}

\makeatletter
\g@addto@macro\normalsize{%
  \abovedisplayskip 2pt plus1pt 
  \belowdisplayskip 2pt plus1pt
  \abovedisplayshortskip  2pt plus1pt%
  \belowdisplayshortskip  1pt plus1pt
}
\makeatother

\copyrightyear{2023}
\acmYear{2023}
\setcopyright{acmlicensed}\acmConference[SIGIR '23]{Proceedings of the 46th International ACM SIGIR Conference on Research and Development in Information Retrieval}{July 23--27, 2023}{Taipei, Taiwan}
\acmBooktitle{Proceedings of the 46th International ACM SIGIR Conference on Research and Development in Information Retrieval (SIGIR '23), July 23--27, 2023, Taipei, Taiwan}
\acmPrice{15.00}
\acmDOI{10.1145/3539618.3591777}
\acmISBN{978-1-4503-9408-6/23/07}

\begin{document}

\title[Topic-oriented Adversarial Attacks against Black-box Neural Ranking Models]{Topic-oriented Adversarial Attacks\\ against Black-box Neural Ranking Models}

\author{Yu-An Liu}
\orcid{0000-0002-9125-5097}
\author{Ruqing Zhang}
\authornote{Research conducted when the author was at the University of Amsterdam.}
\orcid{0000-0003-4294-2541}
\affiliation{
	\institution{CAS Key Lab of Network Data Science and Technology, ICT, CAS}
	\institution{University of Chinese Academy of Sciences}
	\city{Beijing}
	\country{China}
}
\email{{liuyuan21b,zhangruqing}@ict.ac.cn}

\author{Jiafeng Guo}
\orcid{0000-0002-9509-8674}
\authornote{Jiafeng Guo is the corresponding author.}
\affiliation{
	\institution{CAS Key Lab of Network Data Science and Technology, ICT, CAS}
	\institution{University of Chinese Academy of Sciences}
	\city{Beijing}
	\country{China}
}
\email{guojiafeng@ict.ac.cn}

\author{Maarten de Rijke}
\orcid{0000-0002-1086-0202}
\affiliation{
 \institution{University of Amsterdam}
 \city{Amsterdam}
 \country{The Netherlands}
}
\email{m.derijke@uva.nl}

\author{Wei Chen}
\orcid{0000-0002-7438-5180}
\affiliation{
	\institution{CAS Key Lab of Network Data Science and Technology, ICT, CAS}
	\institution{University of Chinese Academy of Sciences}
 \city{Beijing}
 \country{China}
}
\email{chenwei2022@ict.ac.cn}

\author{Yixing Fan}
\orcid{0000-0003-4317-2702}
\affiliation{
	\institution{CAS Key Lab of Network Data Science and Technology, ICT, CAS}
 \institution{University of Chinese Academy of Sciences}
 \city{Beijing}
 \country{China}
}
\email{fanyixing@ict.ac.cn}

\author{Xueqi Cheng}
\orcid{0000-0002-5201-8195}
\affiliation{
	\institution{CAS Key Lab of Network Data Science and Technology, ICT, CAS}
	\institution{University of Chinese Academy of Sciences}
	\city{Beijing}
	\country{China}
}
\email{cxq@ict.ac.cn}

\renewcommand{\shortauthors}{Yu-An Liu et al.}

\begin{abstract}

Neural ranking models (NRMs) have attracted considerable attention in information retrieval. Unfortunately, NRMs may inherit the adversarial vulnerabilities of general neural networks, which might be leveraged by black-hat search engine optimization practitioners. Recently, adversarial attacks against NRMs have been explored in the paired attack setting, generating an adversarial perturbation to a target document for a specific query. In this paper, we focus on a more general type of perturbation and introduce the topic-oriented adversarial ranking attack task against NRMs, which aims to find an imperceptible perturbation that can promote a target document in ranking for a group of queries with the same topic. We define both static and dynamic settings for the task and focus on decision-based black-box attacks. We propose a novel framework to improve topic-oriented attack performance based on a surrogate ranking model. The attack problem is formalized as a Markov decision process (MDP) and addressed using reinforcement learning. Specifically, a topic-oriented reward function guides the policy to find a successful adversarial example that can be promoted in rankings to as many queries as possible in a group. Experimental results demonstrate that the proposed framework can significantly outperform existing attack strategies, and we conclude by re-iterating that there exist potential risks for applying NRMs in the real world.

\end{abstract}

\ccsdesc[500]{Information systems~Adversarial retrieval}

\keywords{Neural ranking model, Adversarial attack, Reinforcement learning}

\maketitle

\vspace*{-2mm}
\section{Introduction}

Ranking models are the main components of information retrieval (IR) systems. 
Building on advances in deep neural networks \cite{lecun2015deep}, neural ranking models (NRMs) \citep{ZhuyunDai2019DeeperTU,guo2016deep,mitra2017learning,KezbanDilekOnal2018NeuralIR} have achieved promising ranking effectiveness. 

\heading{Vulnerability of NRMs} Besides effectiveness, recently, robustness of NRMs has received increasing attention from the research community~\cite{wu2022neural}. 
In many natural language processing (NLP) \cite{JavidEbrahimi2017HotFlipWA, CongzhengSong2020AdversarialSC} and computer vision (CV) \cite{IanGoodfellow2014ExplainingAH,li2021qair} tasks, deep learning-based models have been found vulnerable to adversarial examples that can trigger the misbehavior with human-imperceptible perturbations. 
In the field of IR, NRMs are also likely to inherit adversarial vulnerabilities of general neural networks~\cite{szegedy2014intriguing}, which  raises legitimate concerns about the robustness and trustworthiness of neural IR systems. 
Therefore, there have been initial studies \citep{wu2022prada, liu2022order, wang2022bert} on adversarial attacks against NRMs.
We believe it is important to study potential adversarial attacks against NRMs in IR as they can identify the vulnerability of NRMs before deploying them in real-world applications and support the development of appropriate countermeasures.  

\noindent\textbf{Paired attack.} Early studies on adversarial attacks against NRMs mainly concern adversarial perturbations over a specific pair of query and document. 
That is, such perturbations are capable of fooling NRMs into promoting a target document in the ranking with respect to a specific query. 
For example, by injecting trigger tokens into a document \cite{liu2022order} or replacing important words in a document with synonyms \cite{wu2022prada}, a target document can be promoted significantly in rankings.  
These prior publications have shown that NRMs are vulnerable to imperceptible adversarial perturbations. 

\begin{figure}[t]
    \centering
    \includegraphics[scale=0.56]{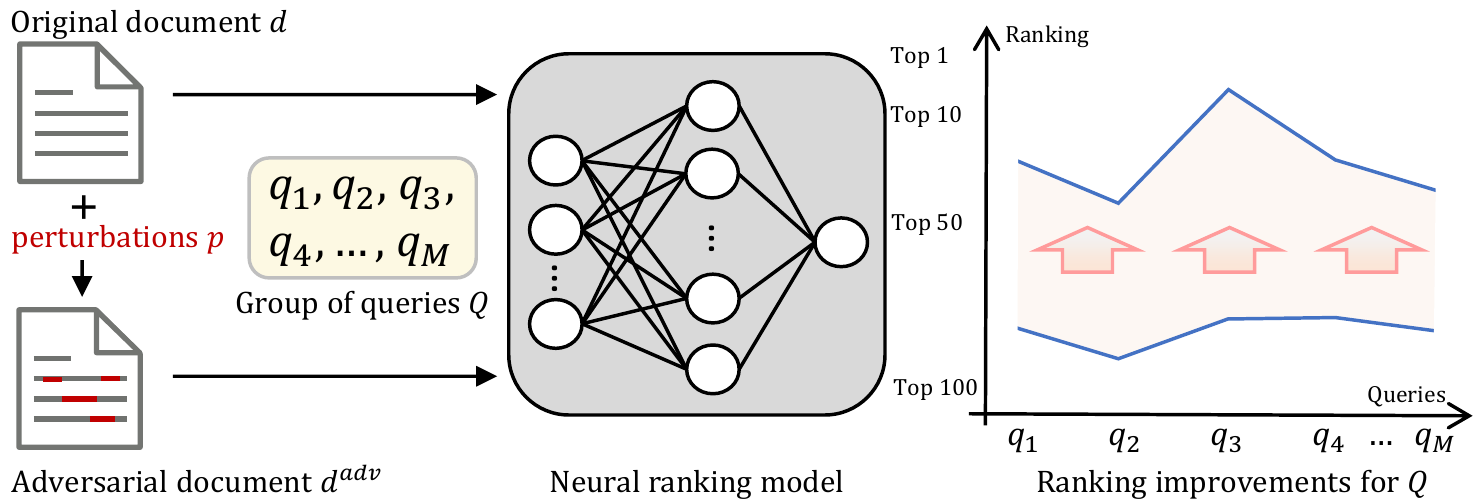}
    \caption{The topic-oriented adversarial ranking attack (TARA) task.}
    \label{fig:TASK STATEMENT}
\end{figure}

\heading{Topic-oriented group attack} In this paper, we introduce a more general attack task, namely the \emph{topic-oriented adversarial ranking attack} (TARA) task against NRMs. 
Given a neural ranking model and a group of queries with the same topic, TARA aims to promote a target document in the rankings with respect to each query in the group, by perturbing the document's text in a semantic-preserving way.  
See Figure~\ref{fig:TASK STATEMENT} for a visualization. 
This attack scenario is more practical than the existing paired attack scenario, since it is more economic for black-hat search engine optimization (SEO) \cite{gyongyi2005web} to exploit document perturbations in terms of query topics rather than in terms of individual queries, to improve campaign performance \cite{kurland2022competitive}.
E.g., in paid search advertising, when advertisers create an advertisement, they select a set of keywords for a group of target queries with the same topic \cite{gyongyi2005web}, e.g., a shoe seller wants clicks from users who intend to buy shoes with different queries with the topic of ``shoe'', like ``shoes for running'' and ``hiking shoe''.   
Compared with paired attacks, the topic-oriented group attack is significantly more challenging as it must consider relationships between a document and an entire group of queries, instead of a single query, so as to find the generic vulnerability of NRMs. 

We focus on a practical and challenging decision-based black-box setting, where no model information is exposed except that the attackers can query the target NRM and obtain hard-label predictions~\citep{wu2022prada,liu2022order}. 
Unlike existing ranking attacks that are only designed for static scenarios, we define both static and dynamic settings for the TARA task, where the target NRM remains unchanged (the static setting) or is kept up-to-date in the presence of potentially highly dynamic corpora (the dynamic setting). 
The dynamic setting is inspired by the use of practical IR systems, as the majority of websites encapsulating online information are dynamic \citep{kurland2022competitive}. 
To facilitate our study and evaluation of the TARA task, we build two benchmark datasets based on several public IR collections, i.e., MS MARCO, ORCAS, ClueWeb09-B, and Web Track in TREC 2012. 

\heading{An RL-based ranking attack method} Typically, an attack process in TARA can be regarded as a series of interactions between the attacker and the target NRM. 
During these interactions, the target NRM may be dynamically updated due to changing corpora and the attacker is expected to generate document perturbations that can fool the current NRM for a group of queries with high probability. 
Therefore, we introduce a novel \emph{reinforcement learning-based adversarial ranking attack framework} (RELEVANT), to generate adversarial examples, which uses a Markov decision process \cite{sutton2018reinforcement} to track interactions between the attacker and the target NRM. 
First, we train a surrogate ranking model that can mimic the ranked list of the black-box target NRM. 
We set the surrogate model as a virtual environment and design several types of environmental dynamics for updating the corpus. 
Then, we explore two attack strategies as the agent to generate document perturbations, which interact with the environment to get information about how good the ranking promotions are. 
To guide the agent to generate a general perturbation for a group of queries, we design a topic-oriented reward by contrasting the relevance between each query and the attacked document at the current state with that at the previous state. 
During the RL process, the attack strategy is updated continuously based on the current NRM's real-time feedback. 

\heading{Experimental findings} 
Experimental results under both static and dynamic scenarios demonstrate the vulnerability of existing NRMs and the effectiveness of our adversarial attack method. 
We also provide detailed analyses and conduct case studies to gain a better understanding of the learned document perturbations. 

\vspace*{-2mm}
\section{Related Work}

\noindent\textbf{Text ranking models}. 
Ranking models have always been at the heart of IR. 
Over the past decades, ranking models have experienced rapid algorithmic shifts, from early heuristic models \cite{GerardSalton1975AVS}, probabilistic models \cite{StephenRobertson1994SomeSE,JayPonte1998ALM} to modern learning to rank models \cite{li2014learning,TieYanLiu2009LearningTR}.  
With the development of deep learning, researchers have adopted NRMs \cite{ZhuyunDai2019DeeperTU,guo2016deep,mitra2017learning,KezbanDilekOnal2018NeuralIR}, which have been proved to be effective in capturing latent semantics and ranking features. 
Recently, researchers have also investigated applying popular pre-trained language models for text ranking \cite{RodrigoNogueira2019PassageRW,IssaAnnamoradnejad2020ColBERTUB}, which achieves new state-of-the-art performance \cite{fan2022pre}. 
Besides direct applications, prior work demonstrates that crafting pre-training objectives tailed for IR \cite{ma2021prop,ma2021pre} can further enhance the performance on downstream ranking tasks. 
However, these text ranking models also inherit the adversarial vulnerabilities of neural networks, which remain under-explored. 

\heading{Adversarial attacks} 
Deep neural networks are notorious for their vulnerability to adversarial examples, which are crafted with imperceptible perturbations to the original input~\cite{szegedy2014intriguing}.
This has motivated research into adversarial attacks \cite{IanGoodfellow2014ExplainingAH,AleksanderMadry2018TowardsDL} to find a minimal perturbation that maximize the model's risk of making wrong predictions. 
Adversarial attacks can be grouped in \emph{white-box}~\cite{JavidEbrahimi2017HotFlipWA} and \emph{black-box} attacks~\cite{papernot2017practical}.  
Adversarial attacks have been explored in natural language processing (NLP) and computer vision (CV) tasks, e.g., text classification \citep{JavidEbrahimi2017HotFlipWA,EricWallace2019UniversalAT}, image classification \citep{IanGoodfellow2014ExplainingAH,AleksanderMadry2018TowardsDL}, and image retrieval \cite{li2021qair,MingyangChen2021DAIRAQ}.  
In IR, search engine optimization (SEO) has been around  since the dawn of the web; white-hat \cite{goren2020ranking} and black-hat \cite{castillo2011adversarial} SEO are distinguished based on whether the intention to modify the document is malicious. 
We focus on adversarial attacks against NRMs, which can be regarded as a new type of black-hat SEO.

There has been limited research regarding this direction.  
E.g., some work \cite{raval2020one,CongzhengSong2020AdversarialSC} explore token perturbations' impact on document ranking and \citet{wang2022bert} investigate BERT-based ranking model attacks. 
However, such work addresses the white-box attack scenario and ignores the practical conditions of invisibility of the target model. 
\citet{wu2022prada} and \citet{liu2022order} propose black-box attacks using word substitution and trigger generation.
Unlike these paired attack methods, we propose to promote a target document in rankings with respect to a group of queries with the same topic. 

As a special attack in NLP and CV, universal adversarial perturbations \cite{moosavi2017universal,EricWallace2019UniversalAT}  have been proposed, where the same attack perturbation can be applied to any input to the target model. Our  attack can be seen as a typical case of universal attacks in IR, i.e., a single document perturbation for a group of queries.

\heading{Reinforcement learning}
Reinforcement learning (RL) \cite{sutton2018reinforcement} is a widely used machine learning approach involving exploration and exploitation. 
It has been successfully applied in various applications \cite{arulkumaran2017deep}, e.g., games \cite{DavidSilver2022MasteringTG}, CV \cite{JuanCCaicedo2015ActiveOL}, NLP \cite{yu2017seqgan}, and IR~\cite{huang-2022-state}. 
Recently, some work has applied RL methods to generate adversarial examples in NLP tasks~\cite{vijayaraghavan2019generating,xu2019lexicalat,zou2020reinforced}.
\citet{maimon2022universal} learn a single search policy over a predefined set of semantics for text classifiers.  
Unlike this work, we aim to generate fluent and semantic-preserving adversarial examples against NRMs by optimizing the evaluation metrics of the TARA task through a deep RL approach. 
To simulate corpus dynamics, we leverage a surrogate ranking model as a virtual environment with several dynamic settings. 

\vspace*{-2mm}
\section{Problem Statement}

We introduce the TARA task and describe the benchmark datasets.

\vspace*{-2mm}
\subsection{Task description}

In ad-hoc retrieval, given a query $q$ and a set of $N$ document candidates $\mathcal{D} = \{d_1,d_2,\ldots,d_N\}$ from a corpus $\mathcal{C}$, a ranking model $f$ aims to associate a relevance score $f(q,d_n)$ with each pair of $q$ and $d_n \in \mathcal{D}$ to rank the whole candidate set. 
For example, the ranking model outputs the ranked list $L = [d_N, d_{N-1},\ldots, d_1]$ if it determines $f(q,d_N) > f(q,d_{N-1})> \cdots > f(q,d_1)$. 

\heading{Objective of the adversary} 
The TARA task is to find an optimized topic-oriented and very small perturbation, which fools the NRMs into promoting the target document in rankings with respect to a group of queries with the same topic.  
Formally, given a target document $d$ and a group of queries with the same topic  $Q=\{q_1,\dots,q_M\}$, the goal is to construct a valid adversarial example $d^{adv}$ that can be ranked higher to each query $q_m \in Q$ by NRMs while resembling $d$.  
We use a soft objective to measure the success of the TARA task.

Specifically, we say $d^{adv}$ succeeds to attack the group of queries $Q$ with level $\alpha \in [0,1]$, if there exists $Q_{\alpha}$ with $|Q_\alpha|/|Q|\geq \alpha$, such that for all $q_m$ in $Q_\alpha$:
\begin{equation}
\label{eq:form}
\operatorname{Rank}(q_m,d^{adv}) < \operatorname{Rank}(q_m,d) \text{ such that }\operatorname{Sim}(d,d^{adv}) \geq \epsilon,
\end{equation}
where $\operatorname{Rank}(q_m,d)$ and $\operatorname{Rank}(q_m,d^{adv})$ denote the position of $d$ and $d^{adv}$ in the ranked list with respect to each query $q_i$, respectively.  
A smaller value of rank position denotes a higher ranking. 
$\operatorname{Sim}: \mathcal{D} \times \mathcal{D} \to (0,1)$ refers to a  similarity function and $\epsilon$ is the minimum similarity between $d$ and $d^{adv}$. 
The adversarial example $d^{adv}$ can be regarded as $d + p$, where $p$ denotes the perturbation to $d$. 
Ideally, $d^{adv}$ should be semantically consistent with $d$ and imperceptible to human judges yet misleading to NRMs.

\noindent\textbf{Decision-based black-box attacks.} 
We focus on decision-based black-box attacks against NRMs for TARA task, because most real-world search engines are black boxes and only provide hard-label outputs. 
The adversary can only query the target NRM to obtain corresponding rank positions of the partially retrieved list \cite{wu2022prada}. 

\heading{Static and dynamic settings} 
An essential characteristic of search engines operating over the web, is its inherently dynamic nature, with the corpus change.  
Though some studies \cite{wu2022prada,liu2022order} have viewed the ranking attack as being interactive, they simply consider the target NRM to be static and learn fixed attack strategies. 

In this work, we define two settings of the target NRM according to its update frequency: 
\begin{enumerate*}[label=(\roman*)]
\item Static: The target NRM is fixed during the attack without continuous update; and 
\item Dynamic: The target NRM is updated  in real-time along with the dynamic of the corpora. 
\end{enumerate*}
The attacker should maintain the attack performance even if the search environment is dynamically updated.  
Although the dynamic setting for the TARA task is significantly more challenging than the static setting, it is more practical and enables broader applicability of the attack methods to a real-world search engine. 

\vspace*{-2mm}
\subsection{Benchmark construction}

To evaluate the TARA task, we build benchmark datasets based on two public collections: 
\begin{enumerate*}[label=(\roman*)]
\item ClueWeb09-B~\cite{clarke2009overview} with 150 queries from TREC Web Tracks 2009-2011 and 50M documents; and
\item MS MARCO Document Ranking (MS MARCO)~\cite{nguyen2016ms} with about 0.37 million training queries and 3.2 million documents.
\end{enumerate*}
We build groups of queries with the same topic as follows. 

\heading{ClueWeb09-B} 
We leverage the TREC 2012 Web Track \cite{clarke2012overview} to construct groups of queries. 
Specifically, the TREC 2012 Web Track selects 50 queries from ClueWeb09-B and every query is structured as a representative group of subtopics, each related to a different user need.
The selection of subtopics attempts to reflect a mix of genuine user search intent. 
Therefore, we directly use these 50 groups of queries of related topics as the query set to perform attacks. 
We leave the remaining 100 queries in the ClueWeb09-B without subtopics to train the target ranking model. 

\begin{table}[t]
\renewcommand{\arraystretch}{0.9}
\setlength\tabcolsep{11.2pt}
  \caption{Data statistics: \#q denotes the number of queries, \#d denotes the number of target documents, and \#w denotes the number of words.}
  \label{tab:Data statistics.}
  \begin{tabular}{lcc}
    \toprule
    & Q-MS MARCO & Q-ClueWeb09 \\
    \midrule
    Group of queries & 200\phantom{.00} & \phantom{0}50\phantom{.00}  \\
    Group: avg \#q & \phantom{0}20\phantom{.00} & \phantom{00}5.84  \\
    Group: avg \#d & \phantom{0}10\phantom{.00} & \phantom{0}10\phantom{.00}  \\ 
    Query: avg \#w & \phantom{00}4.72 & \phantom{00}7.93 \\
    Document: avg \#w & 408.19 & 795.56 \\
  \bottomrule
\end{tabular}
\end{table}

\heading{MS MARCO} 
We leverage the training data to train the target model and leverage the development set with 5,193 queries to construct groups of queries for attacks.  
The number of queries in the development set is insufficient to support the collection of a large number of queries with the same topics.
The Open Resource for Click Analysis in Search (ORCAS) dataset \cite{craswell2020orcas} is a large-scale dataset of click logs related to documents in MS MARCO, with over 10 million distinct queries. 
ORCAS is a supplement to the MS MARCO training set, and the queries in it are distributed across common and rare terms~\cite{craswell2020orcas}. 
We use ORCAS to help aggregate queries in the MS MARCO dataset. 
We randomly select 200 queries from the MS MARCO development set and use Sentence-BERT \cite{reimers2019sentence} to obtain representations of each query in ORCAS and the selected queries in MS MARCO. 
For each selected query in MS MARCO, we use cosine similarity to calculate the top 100 similar queries from ORCAS.  
We then randomly select 19 queries from the top 100 results for diversity, and 20 similar queries with the same topic are grouped. 

For each group of queries, we construct target documents for attacks.  
Following \cite{wu2022prada,wang2022bert}, we attack 10 documents ranked in the top 100 documents. 
Since a document has different rankings for each query in a group of queries, we use the average ranking under all queries in a group to measure the overall document relevance. 
Specifically, we randomly choose 10 marginal documents from each group of queries with an average ranking of 95-100 following \cite{liu2022order}.  
We refer to the benchmark datasets constructed based on MS MARCO and ClueWeb09-B as Q-MS MARCO and Q-ClueWeb09, respectively. 
Table~\ref{tab:Data statistics.} shows the overall statistics. 

\begin{figure}[t]
    \centering
    \includegraphics[width=1\linewidth]{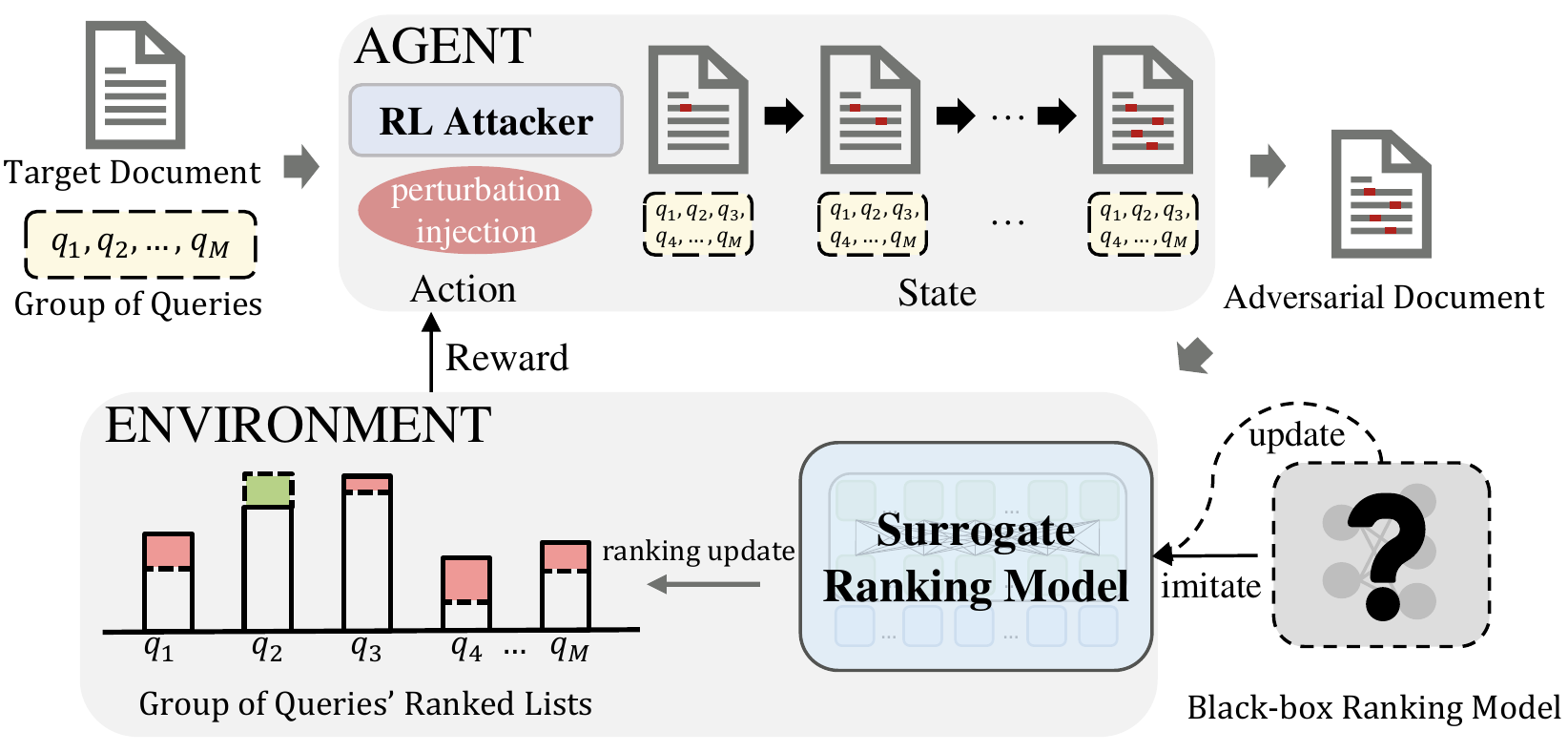}
    \caption{RL-based attack framework (RELEVANT).}
    \label{fig:framework}
\end{figure}

\vspace*{-2mm}
\section{Method}

We introduce RELEVANT, our RL-based attack framework for the TARA task.

\vspace*{-2mm}
\subsection{Motivation}

The attack process in TARA can be regarded as a series of interactions between the attacker and the target NRM: 
the attackers modify the target document, and the NRM ranks the attacked document.  
Then, the attacker observes the rank change to a desired group of queries and further optimizes its attack strategy. 
During a sequence of interactions, the target NRM may dynamically change, and the attack strategies are expected to change accordingly.  

We introduce an RL-based framework RELEVANT to learn an optimal ranking attack strategy that fits the dynamically changing NRM.  
We take the document owner as the agent, their topic-oriented attack perturbations as the action, and treat the target NRM as the environment. 
As shown in Figure~\ref{fig:framework}, the RELEVANT framework consists of two major components: 
\begin{enumerate*}[label=(\roman*)]
\item A surrogate ranking model, which imitates the behavior of the target NRM and serves as the virtual environment. 
\item An RL Attacker, which receives topic-oriented rewards from the environment and generates general document perturbations under the group of queries. 
\end{enumerate*}

\vspace*{-2mm}
\subsection{Environment: Surrogate ranking model}

Under the decision-based black-box setting, we train a surrogate ranking model to imitate and achieve comparable performance to the target NRM as the virtual environment for the RL Attacker. 
We design various corpus dynamics to simulate the ever-changing web. 

\subsubsection{\textbf{Black-box ranking model imitation}}

Following \cite{wu2022prada,liu2022order}, we leverage the relative relevance information among the ranked result list returned by the target model to construct a synthetic dataset, for training a surrogate model. 
Formally, given a query $q$ from a query collection $\mathcal{Q}$ from the downstream search tasks, we get the rank list $L$ of $N$ documents returned by the target NRM.   
We generate pseudo-labels as the ground-truth by treating the top-$k$ ranked documents $L[:k]$ as relevant documents while treating the other documents $L[k+1:N]$ as irrelevant documents.  
We initialize the surrogate ranking model $\tilde{f}$ using the original BERT and the objective function to train $\tilde{f}$ is defined as: 
\begin{equation}
\mathcal{L} = \frac{1}{|\mathcal{Q}|}\sum_{q \in \mathcal{Q}} \max(0, \eta - \tilde{f}(q, L[:k]) + \tilde{f}(q, L[k+1:N])),
\end{equation}
where $\eta$ is the margin for the hinge loss function.

\subsubsection{\textbf{Dynamics of environment}}

Besides the static setting, where the target ranking model is fixed, we also envisage three dynamic cases of the environment given the dynamic nature of the web.

\heading{Document incremental}
In the real world, new documents usually arrive sequentially instead of simultaneously. 
Here, we add new documents in each update round for training the target model in response to incremental information. 

\heading{Document removal}
A document may also be removed from the corpus. Here, we randomly delete some documents from the corpus and re-train the target model. 

\heading{Ranking-incentivized document} 
The search system may detect that a document is promoted significantly in the short term, and then design corresponding countermeasures to force it to be outside the top-ranked list. Here, we simply regard the attacked documents as abnormal and mix them into negative examples for further training the target model. 

In order to ensure the evolution of attacks, the surrogate model needs to be updated along with the target model.
Instead of retraining, we continue to train the surrogate model with another epoch using the sampled data from the target model. 

\vspace*{-2mm}
\subsection{Agent: RL attacker} 
We explore two attack strategies based on word-level and sentence-level textual attacks \cite{zhang2020adversarial}. 
For word-level attacks, we use word substitution \cite{jin2020bert}, the core idea of which is to select the important words in the document for synonym substitution.
For sentence-level attacks, we use the trigger generation \cite{EricWallace2019UniversalAT}, which generates a generic sentence to be injected at the beginning of the document. 

In general, the attack process under the above environment can be regarded as a sequential decision process during which the RL attacker decides the perturbations to the target document. 
Therefore, we mathematically formalize the search process as an MDP, which is described by a tuple $\langle\mathcal{S}, \mathcal{A}, \mathcal{T} , R, \gamma \rangle$ including the state, action, transition, reward, and discount factor. 
Specifically, $\mathcal{S}$ denotes the state space, and $\mathcal{A}$ denotes the action space. 
$\mathcal{T}: \mathcal{S} \times \mathcal{A} \to \mathcal{S}$ is the transition function that generates the next state $s_{t+1}$ from the current state $s_t$ and action $a_t$. 
$R: \mathcal{S} \times \mathcal{A} \to \mathbb{R}$ is the reward function, while the reward at the $t$-th step is $r_t = R(s_t,a_t)$.
$\gamma \in [0,1]$ is the discount factor for future rewards.
Formally, the MDP components are specified with the following definition: 

\begin{itemize}[nosep,leftmargin=*]
\item \textbf{State} $s$ is the document, with an initial state $s_0$ is a target document and a terminal state is a successful adversarial example. 
\item \textbf{Action} $a$ is a perturbation the RL Attacker selects to inject into the document. 
We aim for these actions, i.e., word substitution and trigger generation, to preserve the document's semantic.  
\item \textbf{Transition} $\mathcal{T}$ changes the state of the document $d$, adding one perturbation at each time step.
\item \textbf{Reward} $\mathcal{R}$ is the reward function given by the simulated ranking model to provide supervision signals for the model training. 
\end{itemize}

\noindent%
We solve the MDP problem with the policy gradient algorithm REINFORCE \cite{sutton2018reinforcement}. 
At each time step $t$, the policy $\pi(a_t\mid s_t)$ defines the probability of sampling action $a_t \in \mathcal{A}$ in state $s_t \in \mathcal{S}$. 
The aim of RL is to learn an optimal policy $\pi^*$ by maximizing the expected cumulative reward $R_t = \mathbb{E}[\sum_{k=0}^{\infty} \gamma^k r_{t+k} ]$. 

\vspace*{-1mm}
\subsubsection{\textbf{Topic-oriented reward design}}

A good reward function should gradually guide the agent toward the final topic-oriented goal. 
We define multiple subgoals between the original document and a successful adversarial example to provide positive feedback when each subgoal is achieved. 
We define the anchor document, i.e., the document in the returned list whose ranking position is higher than the perturbed document at the last state to each query in a group, as the subgoal. 
The anchor is dynamically changed as the ranking of the after-attacked document may be updated. 
The reward function should be related to the document's ranking to each query in a group, i.e., a perturbed document should receive more rewards if it is ranked higher than the anchor document. 
However, directly using ranking as a reward is sparse. 
  
We shape the reward using the surrogate model's relevance scores as a potential function. 
Considering the global effect of the attack: 
\begin{enumerate*}[label=(\roman*)]
\item If the target document is not ranked higher than its anchor document for all queries $q \in Q$, the attack fails.
In this case, we use a fixed penalty factor $\xi$ as the reward.
\item Conversely, if the attack succeeds, we use the maximum improvement in relevance scores between the current perturbed document and the anchor document to motivate the agent to learn an effective attack strategy. 
\end{enumerate*}

These assumptions lead us to define the topic-oriented reward function for a group of queries as follows: 
\begin{equation}
R(s_t,a_t) =\left\{
\begin{aligned}
-\xi, \quad \text{if } \min(\{\tilde{f}(q,d_{t})-\tilde{f}(q,d_{t}^{A})\}_{q \in Q}) < 0 \\
\max(\{\tilde{f}(q,d_{t})-\tilde{f}(q,d_{t}^{A})\}_{q \in Q}) + Cons., \quad  \text{else,}  \\
\end{aligned}
\right.
\end{equation}
where $Cons.$ is a naturalness constraints, for trigger generation  $Cons. =  \beta_1 S_{LM} + \beta_2 S_{NSP}$, $S_{LM}$ is a language model score \cite{LiweiSong2021UniversalAA}, and $S_{NSP}$ is a next sentence prediction score \cite{liu2022order}. 
For word substitution, $Cons. =  \beta_3 S_{SS}$, the $S_{SS}$ is the semantic similarity between the original document and the adversarial example measured by the universal sentence encoder (USE)~\cite{cer2018universal}.
$\beta_1$, $\beta_2$, and $\beta_3$ are the hyperparameters that control the semantic consistency, fluency, and semantic similarity, respectively. 
The surrogate ranking model $\tilde{f}$ predicts the relevance score between the query-document pair.
$d_{t}^{A}$ represents the anchor documents at time step $t$.

\vspace*{-1mm}
\subsubsection{\textbf{Policy network}}
\label{policynetwork}
Exploring different document perturbation methods (actions) requires different learning objectives; thus, we have customized the policy network for each action. 

\heading{Trigger generation} We train a policy network that determines the next generated word based on the existing generated sequence to insert at the beginning of the document in turn.  
The generation starts from the first word of trigger.  
Specifically, every action $a_t$ corresponds to the $t$-th trigger word at the time step $t$ to choose, and is initialized with a $[MASK]$ token. 
In total, there are $T$ words generated as the entire trigger.
The trigger word space is the vocabulary of the surrogate ranking model. 

The process proceeds as follows. 
\begin{enumerate*}[label=(\roman*)]
\item We use the surrogate model $\tilde{f}$ to calculate the pairwise loss  
$\mathcal{L}_{R}(q_i, a_t \oplus s_{t-1}; d_{t}^{A})$, 
where $q_i$ is the $i$-th query in $Q$, $d_{t}^{A}$ is the anchor document at $t$, $s_{t-1}$ is the perturbed document at $t-1$  and  $a_t \oplus s_{t-1}$ denotes injecting a new trigger word in the position $t$.  
\item We obtain the average gradient $\boldsymbol{g}_{s_{t-1},i}$ of $\mathcal{L}_{R}(\cdot)$ with respect to $s_{t-1}$. 
\item We calculate dot products between the embedding matrix $\boldsymbol{E}^{\tilde{f}}$ in $\tilde{f}$  and $\boldsymbol{g}_{s_{t-1},i}$ as the state feature of $s_{t}$. 
\item We input the state feature to the multi-layer perception (MLP) \cite{rumelhart1985learning}, and calculate the probability of any action as
\end{enumerate*}
\begin{equation}
\pi(s_t) = \operatorname{softmax}\left(\operatorname{MLP}\left(\sum_{i}^{|Q|}( [\boldsymbol{E}^{\tilde{f}}]^{T} \cdot \boldsymbol{g}_{s_{t-1},i} )\right)\right). 
\end{equation}
The action is sampled using max sampling~\cite{sutton2018reinforcement} to sample the word with highest probability from the trigger word space, as the next trigger word to be added to the trigger, i.e., $a_{t} = \operatorname{max\,sample}(\pi(s_t))$.

\heading{Word substitution} Here, we aim to substitute important words in the target document $d$ with synonyms. 
The action $a_t$ is to select the $t$-th important word in the document to be substituted, at the time step $t$.
In total, $T$ words in the target document are substituted.
Following \cite{wu2022prada}, we find the importance words in the document that have a strong influence on the rankings. 

The process proceeds as follows.
\begin{enumerate*}[label=(\roman*)]
\item For each word $h_k$ in $d$, we compute its importance score by calculating the gradient $\boldsymbol{g}^{h_k}_{s_{t-1},i}$ of the pairwise loss $\mathcal{L}_{R}(q_i, s_{t-1} ; d_{t}^{A})$ of  $\tilde{f}$ with respect to its embedding vector $\boldsymbol{e}^{\tilde{f}}_{h_k}$ in $\tilde{f}$, where $s_{t-1}$ is the perturbed document at $t-1$ .
\item Then the importance score $I_{h_k,i}$ of each word $h_k$ is calculated by $I_{h_k,i} = \left\| \boldsymbol{g}^{h_k}_{s_{t-1},i} \right\|^2_2 $ .
\item Finally, we concatenate the word-level importance score of every word in the target document as the state feature of $s_t$, and calculate the probability of any action as
\end{enumerate*}
\begin{equation}
\mbox{}\hspace*{-2mm}
\pi(s_t) \! = \! \operatorname{softmax}\left(\operatorname{MLP}\left(\sum_{i}^{|Q|}([I_{h_k,i}])\right)\right), k=1,2, \ldots .
\end{equation}
The output of MLP is the probability of each word in the document to be replaced next. 
We use max sampling to sample the important word with highest probability in the document and then substitute it with a synonym.
For synonym replacement, we use the counter-fitted word embedding space \cite{NikolaMrki2016CounterfittingWV} to obtain synonyms.

\vspace*{-2mm}
\subsection{Training with policy gradient}
In each episode, a trajectory $\tau = s_1,a_1,\ldots,s_T,a_T$ is sampled using policy $\pi$.
The episode terminates when the step $t$ reaches the pre-set limit $T$. 
The training objective is to maximize $J(\theta)$ via:
\begin{equation}
\nabla_\theta J(\theta) = \mathbb{E}_{\pi_\theta}[\nabla_\theta \log \pi_\theta R(\tau)]\nonumber.
\end{equation}
The solution can be approximated by a Monte Carlo estimator \cite{LeventeKocsis2006BanditBM}, i.e., 
$
\nabla_\theta J(\theta) \propto \sum_{u=1}^{U}\sum_{t=1}^{T} \nabla_\theta \log \pi_\theta(a_{u,t}\mid s_{u,t})R_{u,t},
$ where $\theta$ denotes the policy network parameters, $U$ is the number of samples, $T$ is the number of steps.

\section{Experimental setup}

In this section, we introduce our experimental settings. 
The datasets and code are available at \url{https://github.com/ict-bigdatalab/TARA}.

\vspace*{-2mm}
\subsection{Models}

\textbf{Baselines.}
We consider two types of adversarial attack baselines: trigger-based methods and word substitution methods.
We take several representative triggers for comparison (with their proposed policies of trigger generation):
\begin{enumerate*}[label=(\roman*)]
\item Trigger-based term spamming (TS$_{Tri})$ \cite{gyongyi2005web} concatenates randomly sampled terms in the group of queries as a trigger and injects it at the beginning of document. 
\item HotFlip \cite{JavidEbrahimi2017HotFlipWA} is a universal text attack method for NLP to find the optimal trigger via model gradient.  
We compute the gradient of the surrogate model through pairwise loss. 
\item PAT \cite{liu2022order} is a gradient-based ranking attack method empowered by the pairwise objective, to generate triggers for NRMs. 
For HotFlip and PAT, we sum up the pairwise loss of the target document to each query in a group of queries to guide trigger generation. 
\end{enumerate*}

As word substitution methods we consider:
\begin{enumerate*}[label=(\roman*)]
\item Substitution-based term spamming (TS$_{Sub})$ \cite{gyongyi2005web} randomly chooses a starting position in the document and replaces successive words with randomly sampled terms in a group of queries.  
\item PRADA \cite{wu2022prada} is a decision-based ranking attack method for NRMs, which finds important words in the target document for replacement. 
We average the importance weights of each word in the document to each query in a group, as the final weight.  
\item Tf-idf simply replaces the top words in the document with the highest tf-idf scores based on the queries in a group with synonyms. 
\end{enumerate*}

\heading{Model variants} 
We implement several variants of RELEVANT (RELE. for short), denoted as
\begin{enumerate*}[label=(\roman*)]
\item RELE.$_{TG}$ uses the trigger generation strategy as the policy network. 
\item RELE.$_{WS}$ uses the word substitution strategy as the policy network. 
\item RELE.$_{TG-NC}$ removes the naturalness constraints in the reward of trigger generation strategy.
\item RELE.$_{WS-NC}$ removes the naturalness constraints in the reward of word substitution strategy. 
\end{enumerate*}

\vspace*{-2mm}
\subsection{Implementation details}

For MS MARCO and ClueWeb09-B, initial retrieval is performed using the Anserini toolkit \cite{PeilinYang2018AnseriniRR} with the BM25 model to obtain the top 100 ranked documents following \cite{wu2022prada}. For the environment, following \cite{wu2022prada, liu2022order}, we choose the BERT \cite{devlin2018bert} model, which takes the concatenated query and document as input and is fine-tuned with the relevance labels in MS MARCO training set for Q-MS MARCO and ClueWeb09-B training set for Q-ClueWeb09, as the target NRM, respectively. 
We use BERT$_{base}$ as the surrogate model with the length $N$ of the returned list as 100 \cite{wu2022prada}. 
We set $k=1$ for MS MARCO and $k=20$ for ClueWeb09, due to the different numbers of relevant documents per query. 
The margin $\eta$ for the hinge loss is set to 1.
The settings of the dynamic environments are:
\begin{enumerate*}[label=(\roman*)]
\item For document incremental, we first use 60\% of the corpus as the initial training data, and  10\% of the corpus documents are continually added to the corpus at each stage.  
\item For document removal, we randomly remove 10\% of documents from the whole corpus. 
\item For ranking-incentivized documents, the documents that are promoted more than 20 rankings at each stage are considered abnormal. 
\end{enumerate*}
In this way, there are 4 update stages for each dynamic setting.

For the agent, the discounting factor $\gamma$ is set to 0.9, and the hyper-parameters of naturalness constraints ($\beta_1$, $\beta_2$, and $\beta_3$) are set to 0.8, 0.1, and 0.1, respectively. 
The total time steps $T$ for trigger generation and word substitution policy to 5 and 50, respectively. 
The policy network's hidden state dimension for both trigger generation and word substitution is 200.
For fair comparison, we maintain the same trigger length and substitution number in all baselines, at 5 and 50, respectively. 
In the reward, given the minimum ranking improvement of the target document to all queries in a group, if it is less than or equal to 0, higher than 1, or higher than 5, the document whose ranking position is 1, 5 or 10 places higher than the perturbed document is selected as the anchor document, respectively.

Since the environment is explicit under our task setting, the training and testing of our method are performed both on the full dataset, which is a reasonable experimental setup \cite{cuccu2019playing}. 
Specifically, when the training process of RL ends, the policy network stops updating while running another epoch on the full dataset as a testing phase to evaluate the performance. 
In future work, we aim to explore the addition of new query groups and new target documents.  

\vspace*{-2mm}
\subsection{Evaluation metrics}

\textbf{Attack performance.} 
We use three automatic metrics:
\begin{enumerate*}[label=(\roman*)]
\item Q-success rate (QSR) $@\alpha$ (\%), which evaluates the percentage of after-attack documents $d^{adv}$ ranked higher than the original documents $d$ for at least $\alpha (\%)$ queries in $Q$. 
\item Average boosted ranks (avg.Boost), which evaluates the average improved rankings for each target document to each query in a group.  
\item Boosted top-$K$ rate (T$K$R) (\%), which evaluates the percentage of after-attack documents that are promoted into top-$K$ to each query in a group. 
\end{enumerate*}
The effectiveness of an adversary is better with a higher value for all these metrics. 

\heading{Naturalness performance}
Here we use the following:
\begin{enumerate*}[label=(\roman*)]
\item Automatic Spamicity detection, which can detect whether target pages are spam or not.  Following \cite{wu2022prada,liu2022order}, we adopt the utility-based term spamicity method \cite{zhou2009osd} to detect the adversarial examples. 
\item Automatic grammar checkers, which calculates the average number of errors in the attack sequences. Specifically, we use two online grammar checkers, i.e., Grammarly~\citep{grammarly} and Chegg Writing~\citep{chegg-writing}, following~\cite{liu2022order}. 
\item Human evaluation, which measures the quality of the attacked documents following the criteria in \cite{wu2022prada}. 
\end{enumerate*}

\begin{table*}[t]
\centering
   \caption{Attack performance under static environment; $\ast$ indicates significant improvements over the best baseline ($p \le 0.05$).}
   \renewcommand{\arraystretch}{0.9}
   \setlength\tabcolsep{3.5pt}
  	\begin{tabular}{l  c c c c c c   c c c c c c }
  \toprule
  \multirow{2}{*}{Method} & \multicolumn{6}{c}{Q-MS MARCO} & \multicolumn{6}{c}{Q-ClueWeb09}  \\ \cmidrule(r){2-7} \cmidrule(r){8-13}
       & QSR@50\% & QSR@75\% & QSR@100\% & avg.Boost & T10R & T5R  & QSR@50\% & QSR@75\% & QSR@100\% & avg.Boost & T10R & T5R \\ 
       \midrule
TS$_{Tri}$  & \phantom{1}94.9 & 87.9 & 33.9 & 32.6 & 4.3 & 1.8 & \phantom{1}93.6 & \phantom{1}78.0 & 50.8 & 23.6 & 2.3 & 1.1\\
HotFlip  & \phantom{1}50.3 & 39.2 & \phantom{1}9.6 & \phantom{1}8.5 & 0.0 & 0.0 & \phantom{1}48.7 & \phantom{1}36.9 & \phantom{1}8.2 & \phantom{1}6.0 & 0.0 & 0.0\\
PAT   & \phantom{1}90.7 & 81.1 & 27.5 & 21.6 & 1.3 & 0.5 & \phantom{1}89.2 & \phantom{1}76.3 & 40.1 & 18.5 & 0.0 & 0.0 \\
RELE.$_{TG}$ & \textbf{100.0}\rlap{$^{\ast}$} & \textbf{93.4}\rlap{$^{\ast}$} & \textbf{48.6}\rlap{$^{\ast}$}& \textbf{36.7}\rlap{$^{\ast}$} & \textbf{6.5}\rlap{$^{\ast}$} & \textbf{3.3}\rlap{$^{\ast}$} & \textbf{100.0}\rlap{$^{\ast}$} & \textbf{100.0}\rlap{$^{\ast}$} & \textbf{70.0}\rlap{$^{\ast}$} & \textbf{32.1}\rlap{$^{\ast}$} & \textbf{4.7}\rlap{$^{\ast}$} & \textbf{2.6}\rlap{$^{\ast}$} \\
\midrule
TS$_{Sub}$  & 88.6 & 74.7 & 18.6 & 18.6 & 0.8 & 0.3 & 89.8 & 75.2 & 46.1 & 16.8 & 0.8 & 0.5\\
Tf-idf  & 41.6 & 32.5 & \phantom{1}5.8 & \phantom{1}6.9 & 0.2 & 0.0 & 49.0 & 36.1 & \phantom{1}6.2 & \phantom{1}4.3 & 0.0 & 0.0\\
PRADA  & 86.0 & 72.4 & 15.3 & 16.9 & 0.4 & 0.3 & 88.8 & 75.3 & 46.0 & 15.6 & 0.3 & 0.1\\
RELE.$_{WS}$ & \textbf{93.6}\rlap{$^{\ast}$} & \textbf{89.2}\rlap{$^{\ast}$} & \textbf{40.1}\rlap{$^{\ast}$} & \textbf{27.8}\rlap{$^{\ast}$} & \textbf{1.5}\rlap{$^{\ast}$} & \textbf{0.6}\rlap{$^{\ast}$} & \textbf{95.1}\rlap{$^{\ast}$} & \textbf{82.6}\rlap{$^{\ast}$} & \textbf{56.4}\rlap{$^{\ast}$} & \textbf{26.5}\rlap{$^{\ast}$} & \textbf{1.4}\rlap{$^{\ast}$} & \textbf{0.4}\rlap{$^{\ast}$}\\
\bottomrule
    \end{tabular}
   \label{table:Baseline}
\end{table*}
\section{Experimental results}

We first compare the attack performance of RELEVANT and baselines in both static and dynamic environments. 
Then, in the static environment, we evaluate the naturalness of adversarial examples, analyze the ranking model imitation performance, and examine the impact of important hyper-parameters in RELEVANT. 

\vspace*{-2mm}
\subsection{Attack evaluation: Static environment}

\begin{table*}[t]

\centering
   \caption{Attack performance in a dynamic environment; $\ast$ indicates significant improvements over the best baseline ($p \le 0.05$).}
   \renewcommand{\arraystretch}{0.9}
   \setlength\tabcolsep{3.5pt}
  	\begin{tabular}{l  c c c c c c   c c c c c c }
  \toprule
  Method & \multicolumn{6}{c}{Q-MS MARCO} & \multicolumn{6}{c}{Q-ClueWeb09}  \\ 
  \cmidrule{1-1}  \cmidrule(r){2-7} \cmidrule(r){8-13}
     \textbf{DI} & QSR@50\% & QSR@75\% & QSR@100\% & avg.Boost & T10R & T5R  & QSR@50\% & QSR@75\% & QSR@100\% & avg.Boost & T10R & T5R \\ 
       \midrule
TS$_{Tri}$  & 92.3 & 85.3 & 31.0 & 30.1 & 3.9 & 1.5 & 91.6 & 76.3 & 48.6 & 21.2 & 1.9 & 0.8\\
HotFlip  & 32.4 & 23.3 & \phantom{1}4.9 & \phantom{1}4.0 & 0.0 & 0.0 & 30.2 & 21.0 & \phantom{1}4.4 & \phantom{1}3.2 & 0.0 & 0.0\\
PAT   & 81.2 & 72.3 & 21.5 & 18.3 & 0.9 & 0.3 & 79.3 & 69.2 & 32.1 & 12.3 & 0.0 & 0.0 \\
RELE.$_{TG}$ & \textbf{97.7}\rlap{$^{\ast}$} & \textbf{92.8}\rlap{$^{\ast}$} & \textbf{37.7}\rlap{$^{\ast}$} & \textbf{34.2}\rlap{$^{\ast}$} & \textbf{3.2}\rlap{$^{\ast}$} & \textbf{2.0}\rlap{$^{\ast}$} & \textbf{96.4}\rlap{$^{\ast}$} & \textbf{86.4}\rlap{$^{\ast}$} & \textbf{66.6}\rlap{$^{\ast}$} & \textbf{29.6}\rlap{$^{\ast}$} & \textbf{2.5}\rlap{$^{\ast}$} & \textbf{1.2}\rlap{$^{\ast}$} \\
\midrule
TS$_{Sub}$  & 86.2 & 72.5 & 16.1 & 16.2 & 0.6 & 0.2 & 88.4 & 73.1 & 45.0 & 15.6 & 0.6 & 0.4\\
Tf-idf  & 39.5 & 30.1 & \phantom{1}5.2 & \phantom{1}6.1 & 0.1 & 0.0 & 47.8 & 35.0 & \phantom{1}5.9 & \phantom{1}4.0 & 0.0 & 0.0\\
PRADA  & 79.8 & 62.5 & 14.2 & 14.0 & 0.3 & 0.0 & 81.2 & 66.2 & 40.4 & 13.7 & 0.1 & 0.0\\
RELE.$_{WS}$ & \textbf{91.7}\rlap{$^{\ast}$} & \textbf{85.3}\rlap{$^{\ast}$} & \textbf{40.9}\rlap{$^{\ast}$} & \textbf{24.3}\rlap{$^{\ast}$} & \textbf{0.9}\rlap{$^{\ast}$} & \textbf{0.3}\rlap{$^{\ast}$} & \textbf{93.0}\rlap{$^{\ast}$} & \textbf{78.2} & \textbf{51.0}\rlap{$^{\ast}$} & \textbf{22.9}\rlap{$^{\ast}$} & \textbf{0.9}\rlap{$^{\ast}$} & \textbf{0.2}\rlap{$^{\ast}$}\\
  \cmidrule{1-1} \cmidrule(r){2-7} \cmidrule(r){8-13}
     \textbf{DR} & QSR@50\% & QSR@75\% & QSR@100\% & avg.Boost & T10R & T5R  & QSR@50\% & QSR@75\% & QSR@100\% & avg.Boost & T10R & T5R \\ 
       \midrule
TS$_{Tri}$  & \textbf{100.0} & \phantom{1}96.9 & 53.5 & 46.2 & 10.1 & 5.8 & \textbf{100.0} & \phantom{1}98.1 & 69.9 & 39.5 & \phantom{1}9.2 & 3.2\\
HotFlip  & \phantom{1}62.1 & \phantom{1}51.6 & 20.5 & 18.0 & \phantom{1}0.8 & 0.3 & \phantom{1}60.2 & \phantom{1}49.8 & 18.9 & 17.2 & \phantom{1}0.2 & 0.0\\
PAT   & \textbf{100.0} & \phantom{1}92.3 & 39.6 & 32.8 & \phantom{1}4.5 & 1.9 & \phantom{1}98.6 & \phantom{1}88.3 & 52.5 & 23.6 & \phantom{1}2.5 & 1.2 \\
RELE.$_{TG}$ & \textbf{100.0} & \textbf{100.0} & \textbf{66.2}\rlap{$^{\ast}$} & \textbf{55.3}\rlap{$^{\ast}$} & \textbf{15.8}\rlap{$^{\ast}$} & \textbf{8.3}\rlap{$^{\ast}$} & \textbf{100.0} & \textbf{100.0} & \textbf{81.4}\rlap{$^{\ast}$} & \textbf{49.6}\rlap{$^{\ast}$} & \textbf{14.2}\rlap{$^{\ast}$} & \textbf{4.7}\rlap{$^{\ast}$} \\
\midrule
TS$_{Sub}$  & \textbf{100.0} & 89.6 & 30.6 & 31.9 & 3.2 & 1.6 & \textbf{100.0} & 90.6 & 60.3 & 28.5 & 2.6 & 1.5\\
Tf-idf  & \phantom{1}56.4 & 43.6 & 10.2 & \phantom{1}8.6 & 0.4 & 0.0 & \phantom{1}48.9 & 39.0 & \phantom{1}9.6 & \phantom{1}6.2 & 0.1 & 0.0\\
PRADA  & \phantom{1}98.8 & 83.6 & 27.3 & 26.9 & 1.6 & 0.7 & \phantom{1}98.3 & 82.3 & 54.0 & 25.1 & 1.4 & 0.6\\
RELE.$_{WS}$ & \textbf{100.0} & \textbf{100.0}\rlap{$^{\ast}$} & \textbf{57.3}\rlap{$^{\ast}$} & \textbf{40.2}\rlap{$^{\ast}$} & \textbf{8.9}\rlap{$^{\ast}$} & \textbf{4.0}\rlap{$^{\ast}$} & \textbf{100.0} & \textbf{93.4} & \textbf{66.2}\rlap{$^{\ast}$} & \textbf{36.0}\rlap{$^{\ast}$} & \textbf{8.0}\rlap{$^{\ast}$} & \textbf{3.1}\rlap{$^{\ast}$}\\
  \cmidrule{1-1} \cmidrule(r){2-7} \cmidrule(r){8-13}
     \textbf{RiD} & QSR@50\% & QSR@75\% & QSR@100\% & avg.Boost & T10R & T5R  & QSR@50\% & QSR@75\% & QSR@100\% & avg.Boost & T10R & T5R \\ 
       \midrule
TS$_{Tri}$  & 50.1 & 38.6 & \phantom{1}9.3 & \phantom{1}7.2 & 0.0 & 0.0 & 46.8 & 38.2 & \phantom{1}8.0 & \phantom{1}5.9 & 0.0 & 0.0\\
HotFlip  & 32.7 & 22.1 & \phantom{1}2.5 & \phantom{1}3.1 & 0.0 & 0.0 & 31.5 & 20.9 & \phantom{1}5.2 & \phantom{1}1.3 & 0.0 & 0.0\\
PAT   & 71.6 & 61.6 & \phantom{1}9.8 & \phantom{1}8.6 & 0.0 & 0.0 & 73.5 & 62.4 & 36.2 & 11.2 & 0.0 & 0.0 \\
RELE.$_{TG}$ & \textbf{80.5}\rlap{$^{\ast}$} & \textbf{71.2}\rlap{$^{\ast}$} & \textbf{19.8}\rlap{$^{\ast}$} & \textbf{17.4}\rlap{$^{\ast}$} & \textbf{0.8}\rlap{$^{\ast}$} & \textbf{0.3}\rlap{$^{\ast}$} & \textbf{81.2}\rlap{$^{\ast}$} & \textbf{68.3}\rlap{$^{\ast}$} & \textbf{30.8}\rlap{$^{\ast}$} & \textbf{10.8}\rlap{$^{\ast}$} & \textbf{0.4}\rlap{$^{\ast}$} & \textbf{0.2}\rlap{$^{\ast}$} \\
\midrule
TS$_{Sub}$  & 41.5 & 35.1 & \phantom{1}6.8 & \phantom{1}5.9 & 0.0 & 0.0 & 42.7 & 22.1 & \phantom{1}6.0 & \phantom{1}5.1 & 0.0 & 0.0 \\
Tf-idf  & 29.8 & 16.7 & \phantom{1}0.8 & \phantom{1}0.6 & 0.0 & 0.0 & 27.5 & 12.1 & \phantom{1}1.0 & \phantom{1}0.3 & 0.0 & 0.0 \\
PRADA  & 66.2 & 49.1 & \phantom{1}6.2 & \phantom{1}5.8 & 0.0 & 0.0 & 68.3 & 55.3 & 21.5 & 10.1 & 0.0 & 0.0 \\
RELE.$_{WS}$ & \textbf{76.4}\rlap{$^{\ast}$} & \textbf{59.2}\rlap{$^{\ast}$} & \textbf{12.8}\rlap{$^{\ast}$} & \textbf{11.0}\rlap{$^{\ast}$} & \textbf{0.3}\rlap{$^{\ast}$} & 0.0 & \textbf{82.0}\rlap{$^{\ast}$} & \textbf{65.4}\rlap{$^{\ast}$} & \textbf{35.8}\rlap{$^{\ast}$} & \textbf{14.0}\rlap{$^{\ast}$} & \textbf{0.6}\rlap{$^{\ast}$} & 0.0\\
\bottomrule
    \end{tabular}
   \label{table:Baseline dynamic}
\end{table*}

Table \ref{table:Baseline} compares the attack performance in a static environment of RELEVANT with trigger generation and word substitution baselines.
We have the following overall observations: 
\begin{enumerate*}[label=(\roman*)]
\item NRMs do inherit adversarial vulnerabilities of deep neural networks and can easily be fooled by the attackers. 
We should, therefore, pay more attention to the potential risks of existing NRMs before deploying them in the real world. 
\item Trigger generation methods generally perform better than word substitution methods. 
The reason may be that trigger generation from the whole vocabulary allows more flexible manipulation than synonym replacement from the document itself. 
Besides, directly adding the trigger may contribute to capturing fine-grained interaction signals between query and document. 
\item The performance of most methods in terms of avg.Boost on Q-ClueWeb09 is lower than that on Q-MS MARCO. 
The reason may be that ClueWeb09's documents come from unprocessed web pages, and the inherent amount of noise could cause the model to be insensitive to the small amount of perturbations added. 
The QSR performance on Q-ClueWeb09 is higher than on Q-MS MARCO. 
There are fewer queries in each group in Q-ClueWeb09, making it easier to take into account the entire group of queries. 
\end{enumerate*}

When we look at the baselines, we find that: 
\begin{enumerate*}[label=(\roman*)]
\item Tf-idf and HotFlip perform poorly, indicating that boosting the document's ranks under a group of queries is non-trivial, which cannot be solved by traditional NLP attack methods or heuristics. 
The customized attack methods for NRMs, i.e., PRADA and PAT, perform better, showing the effectiveness of considering the characteristics of IR.
\item TS$_{Tri}$ and TS$_{Sub}$ perform best among the baselines, showing that it is easy to fool the NRMs by directly using some query terms as a perturbation to the document. 
However, it can be detected by anti-spamming solutions (also observed by \cite{wu2022prada, liu2022order}).  
\end{enumerate*}

Finally, RELEVANT significantly outperforms all baselines in terms of attack performance.
The RL-based framework is helpful by modeling the whole interaction process between the attacker and the target NRM and training with more samples (annotated with rewards).
By leveraging the information of the entire group of queries, RELEVANT outperforms paired attack methods which customize perturbations for a specific query-document pair.

\begin{table}[t]
\centering
   \caption{Attack performance comparisons between the full version of RELEVANT and RELEVANT without naturalness constraints; $\dag$ indicates significant improvements over the full version method ($p \le 0.05$)}
   \renewcommand{\arraystretch}{0.9}
   \setlength\tabcolsep{3pt}
  	\begin{tabular}{l   c c  c c }  \toprule
  \multirow{2}{*}{Method} & \multicolumn{2}{c}{Q-MS MARCO} & \multicolumn{2}{c}{Q-ClueWeb09}  \\ \cmidrule(r){2-3} \cmidrule(r){4-5}
       & QSR@100\% & avg.Boost & QSR@100\% & avg.Boost\\ 
    \midrule
RELE.$_{TG}$ & 48.6 & 36.7 & 70.0 & 32.1 \\
RELE.$_{TG-NC}$ & \textbf{55.8}\rlap{$^{\dag}$} & \textbf{39.2}\rlap{$^{\dag}$} & \textbf{75.3}\rlap{$^{\dag}$} & \textbf{36.2}\rlap{$^{\dag}$} \\
RELE.$_{WS}$ & 40.1 & 27.8 & 56.4 & 26.5 \\
RELE.$_{WS-NC}$ & \textbf{45.1}\rlap{$^{\dag}$} & \textbf{33.7}\rlap{$^{\dag}$} & \textbf{62.9}\rlap{$^{\dag}$} & \textbf{30.3}\rlap{$^{\dag}$} \\
\bottomrule
    \end{tabular}
   \label{table: without constraints}
\end{table}

\vspace*{-2mm}
\subsection{Attack evaluation: Dynamic environment}

The attack performance under dynamic environments is shown in Table \ref{table:Baseline dynamic}. 
The attack baselines are designed for static environments, and we continue updating their attack strategies along with the dynamics of environments. 
Term Spamming and Tf-idf are one-time methods, and the adversarial examples do not change with the environment. 
In three dynamic settings, the model performance after each update stage has a consistent change trend. 
Due to space limitations, we only show the performance after the last ($4$-th) stage. 

As we can see: 
\begin{enumerate*}[label=(\roman*)]
\item For document incremental (DI), the attack success rate of each method is reduced, indicating that the addition of new documents makes the ranking competition more intense. 
\item For document removal (DR), as the number of relevant documents may decrease, the ranking of adversarial documents generated by most attack methods can be improved accordingly. 
\item For ranking-incentivized document (RiD), the performance of all attack methods decreases as the target model's ability to identify anomalous documents increases.
For the one-time attack methods (TS$_{Tri}$, TS$_{Sub}$ and Tf-idf) it is challenging to achieve high rankings compared with static environments, indicating that the spamming-based attack method easily looses its effectiveness once it is struck. 
\item The dynamic environment affects the attack performance to varying degrees. RELEVANT still performs best; it adapts to changes while maintaining the effectiveness of the attack as it can keep trying out new judgments about vulnerabilities of NRMs through the interaction, which brings better generalizability. 
Re-training the attack method from scratch every time the corpus is updated, could incur prohibitively high computational costs. 
The proposed RL-based framework avoids these costs. 
\end{enumerate*}

\vspace*{-2.5mm}
\subsection{Naturalness evaluation}
\label{Naturalness evaluation}
\textbf{Attack performance without naturalness constraints}. 
Table~\ref{table: without constraints} demonstrates the attack performance of the full version RELEVANT and RELEVANT without naturalness constraints.
Although only QSR@100\% and avg.Boost are displayed, the trend is consistent across all evaluation metrics.
Removing the naturalness constraints of RELEVANT enhances attack effectiveness but may increase detection risk. 
Imperceptibility of perturbations on Q-MS MARCO is discussed below, with similar findings on Q-ClueWeb09.

\heading{Automatic spamicity detection} 
Table~\ref{table:anti-spamming} lists the automatic spamicity detection results of RELEVANT and baselines. 
If an example's spamicity score is higher than a detection threshold, it is detected as suspected spam content.
We observe that: 
\begin{enumerate*}[label=(\roman*)]
\item As the threshold decreases, the detector becomes stricter and the detection rate increases for all methods. 
\item Term spamming can be very easily detected since it incorporates many query terms into documents. 
\item The trigger generation methods have a lower upper bound of detection rate, due to the use of fewer words for perturbations.
\item The full version of RELEVANT outperforms the baselines significantly (p-value $< 0.05$), indicating that RELEVANT with naturalness constraints helps adversarial documents to avoid suspicion.
\end{enumerate*}

\begin{table}[t]
\centering
   \caption{The detection rate (\%) via a representative anti-spamming method on the Q-MS MARCO.}
   \renewcommand{\arraystretch}{0.9}
   \setlength\tabcolsep{11pt}
  	\begin{tabular}{l c c c c}
  \toprule
       Threshold & 0.08 & 0.06 & 0.04 & 0.02 \\ 
       \midrule
TS$_{Tri}$  & 24.7 & 35.9 & 56.3 & 80.2 \\
PAT   & \phantom{1}8.0 & 14.5 & 25.3 & 46.2 \\
RELE.$_{TG-NC}$ & \phantom{1}9.7 & 16.9 & 30.2 & 54.5 \\
RELE.$_{TG}$ & \textbf{\phantom{1}6.4} & \textbf{12.2} & \textbf{20.8} & \textbf{38.3} \\
\midrule
TS$_{Sub}$  & 67.6 & 78.3 & 88.9 & 97.1 \\
PRADA   & 11.2 & 18.4 & 30.5 & 50.5 \\
RELE.$_{WS-NC}$ & \phantom{1}9.0 & 14.6 & 24.6 & 46.0 \\
RELE.$_{WS}$ & \textbf{\phantom{1}4.5} & \textbf{\phantom{1}8.0} & \textbf{14.6} & \textbf{25.6} \\
\bottomrule
    \end{tabular}
   \label{table:anti-spamming}
\end{table}

\begin{table}[t]
\centering
   \caption{Online grammar checkers and human evaluation results on the Q-MS MARCO.}
   \renewcommand{\arraystretch}{0.9}
   \setlength\tabcolsep{1pt}
  	\begin{tabular}{l @{} c c   c c   c c}
  \toprule
       Method & Chegg. & Grammar. & Impercept. & \textit{kappa} & Fluency & \textit{Kendall} \\
       \midrule
Original  & 30 & \phantom{1}56 & 0.89 & 0.53 & 4.68 & 0.63\\
\midrule
TS$_{Tri}$  & 42 & \phantom{1}85 & 0.05 & 0.68 & 2.35 & 0.82\\
PAT   & 33 & \phantom{1}65 & 0.73 & 0.46 & 3.85 & 0.91\\
RELE.$_{TG}$ & 32 & \phantom{1}63 & 0.82 & 0.50 & 4.21 & 0.71\\
RELE.$_{TG-NC}$ & 37 & \phantom{1}76 & 0.14 & 0.59 & 2.89 & 0.85\\
\midrule
TS$_{Sub}$  & 62 & 111 & 0.04 & 0.65 & 1.86 & 0.79\\
PRADA   & 53 & \phantom{1}97 & 0.62 & 0.42 & 3.56 & 0.92\\
RELE.$_{WS}$ & 39 & \phantom{1}73 & 0.75 & 0.48 & 4.13 & 0.73\\
RELE.$_{WS-NC}$ & 59 & 107 & 0.34 & 0.53 & 3.16 & 0.95\\
\bottomrule
    \end{tabular}
   \label{table:human evaluation}
\end{table}

\begin{table*}[t]
\centering
   \caption{Attack performance comparisons of RELEVANT between the black-box and the white-box attack setting.}
   \renewcommand{\arraystretch}{0.90}
   \setlength\tabcolsep{2.5pt}
  	\begin{tabular}{l  c c c c c c   c c c c c c }  \toprule
  \multirow{2}{*}{Method} & \multicolumn{6}{c}{Q-MS MARCO} & \multicolumn{6}{c}{Q-ClueWeb09}  \\ \cmidrule(r){2-7} \cmidrule(r){8-13}
       & QSR@50\% & QSR@75\% & QSR@100\% & avg.Boost & T10R & T5R  & QSR@50\% & QSR@75\% & QSR@100\% & avg.Boost & T10R & T5R \\ 
    \midrule
RELE.$_{TG}$ & 100.0 & 93.4 & 48.6 & 36.7 & 6.5 & 3.3 & 100.0 & 100.0 & 70.0 & 32.1 & 4.7 & 2.6 \\
White-RELE.$_{TG}$ & 100.0 & 94.1 & 49.8 & 37.2 & 6.9 & 3.5 & 100.0 & 100.0 & 69.5 & 32.2 & 4.5 & 2.5\\
RELE.$_{WS}$ & \phantom{1}93.6 & 89.2 & 40.1 & 27.8 & 1.5 & 0.6 & \phantom{1}95.1 & \phantom{1}82.6 & 56.4 & 26.5 & 2.4 & 1.4\\
White-RELE.$_{WS}$ & \phantom{1}94.9 & 90.6 & 40.8 & 28.4 & 1.8 & 1.0 & \phantom{1}95.6 & \phantom{1}83.0 & 56.2 & 26.5 & 2.6 & 1.6\\
\bottomrule
    \end{tabular}
   \label{table: black v.s. white}
\end{table*}

\begin{table*}[t]
\centering
   \caption{Attack performance of different hyper-parameter settings on the Q-MS MARCO. The QSR denotes QSR@100\%. }
   \renewcommand{\arraystretch}{0.90}
   \setlength\tabcolsep{4.4pt}
  	\begin{tabular}{l  c c  c c  c c  l  c c  c c  c c }
  \toprule 
 \multirow{3}{*}{Method} & \multicolumn{6}{c}{Position of trigger injection} & \multirow{3}{*}{Method} & \multicolumn{6}{c}{Number of substituted words}  \\ \cmidrule(r){2-7} \cmidrule(r){9-14}
  & \multicolumn{2}{c}{Beginning} & \multicolumn{2}{c}{Middle} & \multicolumn{2}{c}{End} &  & \multicolumn{2}{c}{50} & \multicolumn{2}{c}{30} & \multicolumn{2}{c}{10} 
  \\ \cmidrule(r){2-3} \cmidrule(r){4-5} \cmidrule(r){6-7}  \cmidrule(r){9-10} \cmidrule(r){11-12} \cmidrule(r){13-14}
       & QSR & avg.Boost & QSR & avg.Boost & QSR & avg.Boost & 
       & QSR & avg.Boost & QSR & avg.Boost & QSR & avg.Boost \\ 
       \midrule
PAT & 27.5 & 21.6 & 18.9 & 14.2 & 16.5 & 13.2 & PRADA & 15.3 & 16.9 & 10.5 & \phantom{1}8.2 & \phantom{1}4.6 & \phantom{1}4.1  \\
RELE.$_{TG}$ & 48.6 & 36.7 & 27.5 & 25.3 & 25.2 & 23.3 & RELE.$_{WS}$ & 40.1 & 27.8 & 31.3 & 21.4 & 12.5 & 11.1 \\
\bottomrule
    \end{tabular}
   \label{table: hyper parameter}
\end{table*}

\heading{Automatic grammar checker and human evaluation} 
Table \ref{table:human evaluation} lists the results of the automatic grammar checker and human evaluation, including the annotation consistency test results (the \textit{Kappa} value and \textit{Kendall's Tau} coefficient). 
For human evaluation, we recruit three annotators to annotate 50 randomly sampled adversarial examples and the corresponding documents of each attack method \cite{liu2022order}. 
Following \cite{wu2022prada}, annotators score the \emph{Fluency} of the mixed examples from 1 to 5; higher scores indicate a more fluent examples.  
For \emph{Imperceptibility}, annotators judge whether an example is attacked (labeled as 0) or not (labeled as 1). 
We observe that: 
\begin{enumerate*}[label=(\roman*)]
\item Trigger generation methods generally achieve better fluency and are not easily detected by annotators than word substitution methods. 
\item The Term Spamming method performs poorly under the naturalness metrics, due to the semantic irrelevance between the query terms and the document. 
\item Attack methods with naturalness constraint (i.e., PAT, PRADA, RELEVANT) lag behind the original samples, which indicates that it is not easy to make the attack examples invisible.
Although there is still a gap between the original document and its adversarial example, RELEVANT achieves the best naturalness performance, indicating that the naturalness constraints help generate a natural-looking trigger or synonym.
\end{enumerate*}

\heading{Example triggers}  We randomly sample a group of queries from Q-MS MARCO, in which the query (ID=419333) from MS MARCO is  ``is nizuc resort all inclusive'', and the keywords for its group is ``all inclusive resorts''. 
The target document (ID=D366143) is a resort hotel page, whose average rank to all the queries in the group is $98.5$.
The document begins with: ``about the Pearl South Padre "Skip to main content Account Sign in to see exclusive Member Discount ...''
The trigger generated by our RELEVANT$_{TG}$ and PAT is ``all inclusive resort was'' and ``inclusiveb resortrao taxi all'', respectively. 
By adding the trigger generated by our RELEVANT$_{TG}$ at the beginning of the document, the average rank of the adversarial document is higher than that generated by PAT, i.e.,  $18.3$ vs.\ $31.6$. 
Besides, the trigger generated by RELEVANT$_{TG}$ is more natural-looking and consistent with the document than that generated by PAT.

\vspace*{-2mm}
\subsection{Black-box vs.\ White-box attack}

In this work, we focus on the decision-based black-box attack setting because it is close to the real-world search scenario. 
It is also meaningful to explore the white-box setting to further understand the ranking model's robustness against the TARA. 
First, we evaluate the ranking performance of the surrogate model and the target model over all the queries on the dev sets of the MS MARCO and ClueWeb09-B, respectively. 
The MRR$@10$ of the target and surrogate model on the MS MARCO is $38.61$ and $35.40$, respectively. 
The nDCG$@20$ of the target and surrogate model on the ClueWeb09-B is $27.53$ and $24.95$, respectively.  

Then, to conduct white-box TARA, we directly set the surrogate model as the target NRM and keep other components the same in our RELEVANT$_{TG}$ and RELEVANT$_{WS}$, for which we write White-RELE.$_{TG}$ and White-RELE.$_{WS}$, respectively. 
The results are shown in Table \ref{table: black v.s. white}. 
Even though the white-box setting has full access to the target ranking model, the black-box attack achieves similar performance.
This result shows that the surrogate model is sufficient to mimic the behavior of the target model, which provides the conditions for the transformation of the attack effect of the adversarial examples to the target model.

\vspace*{-2mm}
\subsection{Hyper-parameter sensitivity}

We evaluate RELEVANT with different hyper-parameter settings to investigate how they affect the attack performance on the Q-MS MARCO dataset. The results are shown in Table \ref{table: hyper parameter}. 

We first consider the position of trigger injection. For RELEV-\\ANT$_{TG}$, we to insert triggers at different positions, i.e., the documents' Beginning, Middle, and End.  
We observe that inserting the trigger at the document's beginning achieves the best performance, indicating that the information contained at the beginning of the document matters for interacting with the query.
Next, we consider the number of substituted words. For RELEVANT$_{WS}$, we  substitute different numbers of words (i.e., $10$, $30$, and $50$) in the document. 
We observe that the attack performance of PRADA and RELEVANT$_{WS}$ gradually increase with the increase of the number of substituted words, respectively.  
However, adding the triggers at the beginning of the target document or substituting more words may lead to the attack easily being detected. 
In future work, we will explore more flexible ways to achieve the balance between attack performance and the imperceptibility of adversarial perturbations.

\vspace*{-0mm}
\section{Conclusion and future work}

In this work, we proposed a challenging TARA task against black-box NRMs under both static and dynamic environments, and showed the existence of small general perturbations that can promote the target document in rankings with respect to a group of queries with the same topic.   
We developed an RL-based framework RELEVANT to track the attacker's interactive attack process and continuously update the attack strategies based on the topic-oriented rewards.  
The proposed method along with extensive experiment results reveal the vulnerability and risk of black-box text ranking systems. 

In future work, we would like to explore to adaptively determine the level (character, word, and sentence) of adversarial perturbations for various scenarios and target documents in RELEVANT. 
Beyond the TARA task, the universal adversarial ranking attacks to discover input-agnostic perturbations against NRMs appears to be a promising future direction.

\vspace*{-0mm}
\begin{acks}
This work was funded by the National Natural Science Foundation of China (NSFC) under Grants No. 62006218 and 61902381, the China Scholarship Council under Grants No. 202104910234, the Youth Innovation Promotion Association CAS under Grants No. 20144310 and 2021100, the CAS Project for Young Scientists in Basic Research under Grant No. YSBR-034, the Innovation Project of ICT CAS under Grants No. E261090, the Young Elite Scientist Sponsorship Program by CAST under Grants No. YESS20200121, and the Lenovo-CAS Joint Lab Youth Scientist Project. 
This work was also (partially) funded by the Hybrid Intelligence Center, a 10-year program funded by the Dutch Ministry of Education, Culture and Science through the Dutch Research Council, \url{https://hybrid-intelligence-centre.nl}. 
All content represents the opinion of the authors, which is not necessarily shared or endorsed by their respective employers and/or sponsors.
\end{acks}

\clearpage
\bibliographystyle{ACM-Reference-Format}
\balance
\bibliography{references}

\end{document}